\documentclass[aps,pre,preprint,groupedaddress]{revtex4-1}
\usepackage{chemfig}
\usepackage{geometry}           
\usepackage{graphicx}
\usepackage{amssymb}
\usepackage{epstopdf}
\usepackage{overpic}                                       

\begin{document}
\title{Droplet Breakup of the Nematic Liquid Crystal MBBA}
\author{Benjamin Nachman}
\email[]{bpn7@cornell.edu}
\author{Itai Cohen}
\affiliation{Department of Physics, Cornell University, Ithaca, New York, 14853, USA.}
\date{\today}


\begin{abstract}
Droplet breakup is a well studied phenomena in Newtonian fluids. One property of this behavior is that, independent of initial conditions, the minimum radius exhibits power law scaling with the time left to breakup $\tau$. Because they have additional structure and shear dependent viscosity, liquid crystals pose an interesting complication to such studies. Here, we investigate the breakup of a synthetic nematic liquid crystal (N-(4-{M}ethoxy{b}enzylidene)-4-{b}utyl{a}niline), known as MBBA.  We determine the phase of the solution by using a cross polarizer setup in situ with the liquid bridge breakup apparatus. Consistent with previous studies of scaling behavior in viscous-inertial fluid breakup, when MBBA is in the isotropic phase, the minimum radius decreases as $\tau^{1.03 \pm 0.04}$.  In the nematic phase however, we observe very different thinning behavior. Our measurements of the thinning profile are consistent with two interpretations. In the first interpretation, the breakup is universal and consists of two different regimes. The first regime is characterized by a symmetric profile with a single minimum whose radius decreases as $\tau^{1.51 \pm 0.06}$. The second and final regime is characterized by two minima whose radii decrease as $\tau^{0.52 \pm 0.11}$.  These results are in excellent agreement with previous measurements of breakup in the nematic phase of liquid crystal 8CB and 5CB.  Interestingly, we find that the entire thinning behavior can also be fit with an exponential decay such that $R_{min}\sim \exp\Big((1.2\times 10^2\text{ Hz}) \tau\Big)$. This dependence is more reminiscent of breakup in polymers where entropic stretching slows the thinning process. An analogous mechanism for slowing in liquid crystals could arise from the role played by topological constraints governing defect dynamics. Consistent with this interpretation, crossed polarizer images indicate that significant alignment of the liquid crystal domains occurs during breakup.  
\end{abstract}

\maketitle

\section{Introduction}
One well-studied property of Newtonian fluids is their behavior under droplet breakup.  As a thread of fluid thins until the point of breaking, the minimum thread radius decreases to zero in finite time so that the induced pressure diverges. The balance between the gradient of this divergent pressure and terms in the Navier-Stokes equation that match this divergence determine the limiting behavior of the fluid. The various stress balances that can be acheived in Newtonian fluids have been well studied both in experiments and theory~\cite{Eggers1993,Eggers2008,Suryo2006,Chen2002,Cohen2001,Lister1998,Zhang1999,Day1998,Papagerogiou1995,Cohen1999,Doshi2003,Brenner1996,Keim2006,Brenner1994,Henderson1997,Shi1994,Brenner1997,Eggers1997,Keller1983,Ottino1991}.

Universality in droplet breakup is not a property of all fluids for any set of environmental and initial conditions.  For example, it has been shown that low viscosity drops breaking inside a viscous fluid, such as water inside viscous oil~\cite{Doshi2003} or air inside water~\cite{Keim2006,Schmidt2009,Keim2011}, deviate from universal behavior.  In such cases, the initial and boundary conditions set before breakup are encoded in the asymptotic behavior.  Deviations from universality have also been studied in non-Newtonian fluids.   In particular, polymeric systems have been shown to break slower than a power law fluid~\cite{Amarouchene2001,Roche2011}.   Exponential models seem to fit the data well in these systems where elastic stresses effectively increase the extensional viscosity~\cite{Eggers2008}.

More recently scientists have begun to explore how liquid crystal solutions break. These systems present an additional challenge  since the resulting dynamics arise from coarse grained averaging of small scale flows that are intimately related to the liquid crystalline structure~\cite{Porter2012,Savage2010,Lekkerkerker2012,Zhou2010,Cheong2004,Goyal2008}. The details of how these small scale structures lead to the bulk flows are poorly understood in part because it has been difficult to image the evolution of the liquid crystalline domains during breakup. 

Recent studies have shown that universal behavior can describe breakup dynamics in liquid crystals, but the exact form depends on the phase~\cite{Porter2012,Savage2010,Lekkerkerker2012}.  In particular, for liquid crystals where the domain size is on the same order as the droplet dimensions,  it was found that thinning of the minimum droplet radius depends not only on the phase but specifically on the orientation of the nematic directors inside the fluid~\cite{Zhou2010,Cheong2004,Goyal2008,Lekkerkerker2012}.  Furthermore, it was shown that this thinning can be described in a universal framework in terms of a power law dependance of the minimum radius $R_{min}$ as a function of time left until breakup $\tau$.

Studies have also been conduced for liquid crystals in which the characteristic domain size is much smaller than the drop radius~\cite{Porter2012,Savage2010}.  For example, studies conducted with the thermotropic liquid crystal 8CB (4-cyano 4-octylbiphenyl) and the non-ionic surfactant $C_{12}E_6$ (hexaethylene glycol monododecyl ether) in the {\it smectic} phase, showed universal behavior that could be described by a similarity solution for a strain thinning power-law fluid \cite{Bassaran2006,Bassaran2003,Renardy2002,Renardy2004}.  There, it was shown that the entire breakup was symmetric with a single minimum whose radius $R_{min}\propto \tau^{0.6}$.  Using a cross polarizer setup for the $C_{12}E_6$ studies, it was observed that during breakup the domains within the smectic liquid crystal became more aligned and coarsened - a well established mechanism for strain thinning. Moreover, it was shown that the universal behavior persisted until the neck size was comparable to the liquid crystal domain length scale, which could be set by the sample preparation~\cite{Savage2010}.  

Studies of breakup in 8CB and 5CB (4-cyano 4pentylbiphenl) in the nematic phase showed 
two different breakup regimes~\cite{Porter2012}. Initially the breakup was symmetric and chracterized by $R_{min}\sim \tau^{1.5}$ \cite{Savage2010,Porter2012}. As breakup proceeded, the minimum bifurcated, and the profiles became asymmetric with $R_{min}\sim \tau^{0.7}$. These exponents also showed a slight dependence on temperature. When these liquids were heated so that they underwent a transition into the isotropic phase, it was found that breakup proceeded similarly to Newtonian fluids in the inertial-viscous regime~\cite{Shi1994} where both the initial symmetric and final asymmetric profiles are described by $R_{min}\sim \tau^{1.0}$~\cite{Porter2012}.

The drawback in the 5CB and 8CB samples is that they are opaque and so the cross polarizer technique of~\cite{Savage2010} used to to study $C_{12}E_6$ has never been used to study nematic liquid crystal breakup.  In addition, the thin necks that characterize the final nematic regime make collapsing the profiles onto a universal curve prohibitive.  In the present study, the nematic liquid crystal MBBA (N-(4-{M}ethoxy{b}enzylidene)-4-{b}utyl{a}niline) is used to combat the first of these challenges, namely the inability to use cross polarizers to probe domain orientation.   In addition to studying domain orientation, we determine the phase of the solution with the cross polarizer setup in situ with the liquid bridge breakup apparatus. We determine the minimum radius dependence on phase and analyze the emergent asymptotic behaviors.  

\section{Sample}

\begin{center}
 \begin{figure}[h!]
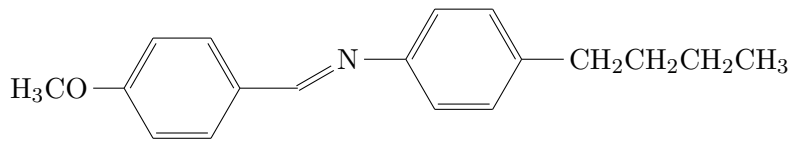

\setatomsep{2em}
\chemfig{
  H_3C{O}-[:0]
      *6(-()=()-(-=[:30]{N}-*6(-()=()-(-[:0]{CH_2CH_2CH_2CH_3})=()-=))=()-=)}
      \caption{Chemical structure of MBBA.}
      \label{fig:MBBA}
\end{figure}
\end{center}

We use MBBA ($98\%$ Sigma-Aldrich) a translucent yellow thermotropic liquid crystal with rod-like molecules that have an aspect ratio of about $2.5$~\cite{viscosity3} (Fig.~\ref{fig:MBBA}). At room temperature it has a density of 1.03g/cc and a surface tension of about $30$ $\mathrm{dyn/cm}$~\cite{SurfaceTension}. Additionally, it has been observed that the surface tension parallel to the direction of alignment is greater by about $8$ $\mathrm{dyn/cm}$ than the surface tension perpendicular to the alignment~\cite{SurfaceTension}. There are many characterizations of the viscosity of a nematic fluid and a complete description for MBBA has been determined in the literature~\cite{viscosity2}.  For example, the Miesowicz coefficient $\eta_3$, which corresponds to the usual shear viscosity of an isotropic fluid~\cite{viscosity3} is about 0.04 $\mathrm{Pa\cdot s}$~\cite{Viscosity} for MBBA at room temperature or roughly $40$ times that of water. 

  Under vacuum, at about $20^\circ$ $\mathrm{C}$, MBBA melts from a crystalline solid into a nematic liquid crystal.  Then, at about $50^\circ$ $\mathrm{C}$, the sample melts into an isotropic fluid~\cite{phasediagram}.  Unfortunately MBBA is very sensitive to the presence of water and will hydrolyze into $p$-anisaldehyde (pAde) and $p-n$ butylaniline (pBa)~\cite{TransitionTemp}, which alters its transition temperatures. In fact this property prevented it from being widely used in LCD screens~\cite{History}.  By maintaining a constant temperature and humidity in the lab we found that the transition temperatures could be stabilized so that the complete phase behavior of MBBA was accessible.  

\begin{figure}
\includegraphics[scale=1.1]{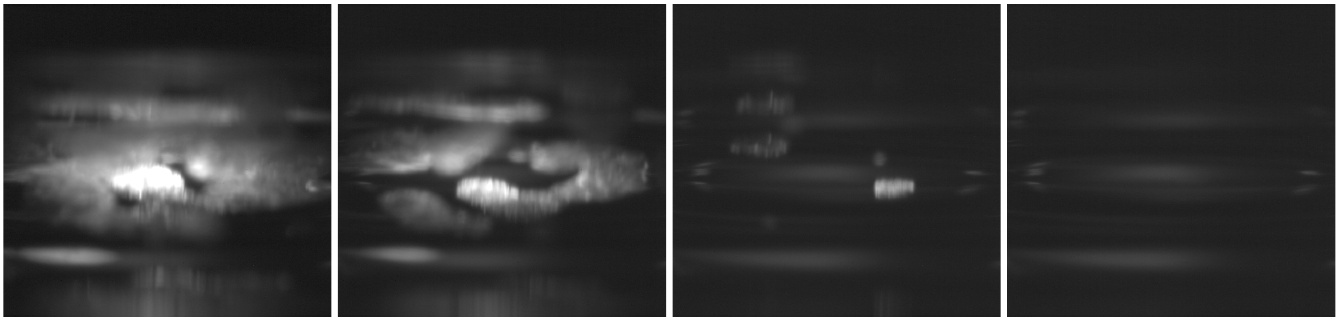}
\caption{Phase transition of MBBA under a cross polarizer.  Below $\sim 20^{\circ}C$, MBBA is a solid. In this series of images, the sample begins melted in the Nematic phase (left image) and transitions into an isotropic phase (right image).  Note that the exact melting point is difficult to identify because it is very sensitive to the level of water in the air.  }
\label{transition}
\end{figure}

Since MBBA is translucent it is possible to probe the phase behavior using cross polarizers.  For example, Fig.~\ref{transition} shows a sample of MBBA under cross polarizers during its phase transition from a nematic to an isotropic liquid at $35^\circ$ $\mathrm{C}$ (Supplementary Movie 1).  Here the nematic portions of the sample appear bright while the isotropic portions are opaque. In the final frame on the right, the backlight can not be seen as the liquid is now in an isotropic phase.  Over all, these in-situ observations provide a new handle for studying breakup in nematic liquid crystals since they enable accurate determination of the phase boundaries associated with the measured breakup dynamics under laboratory conditions. 

\section{Setup and Procedure}
\begin{figure}
\begin{center}
\includegraphics[scale=0.5]{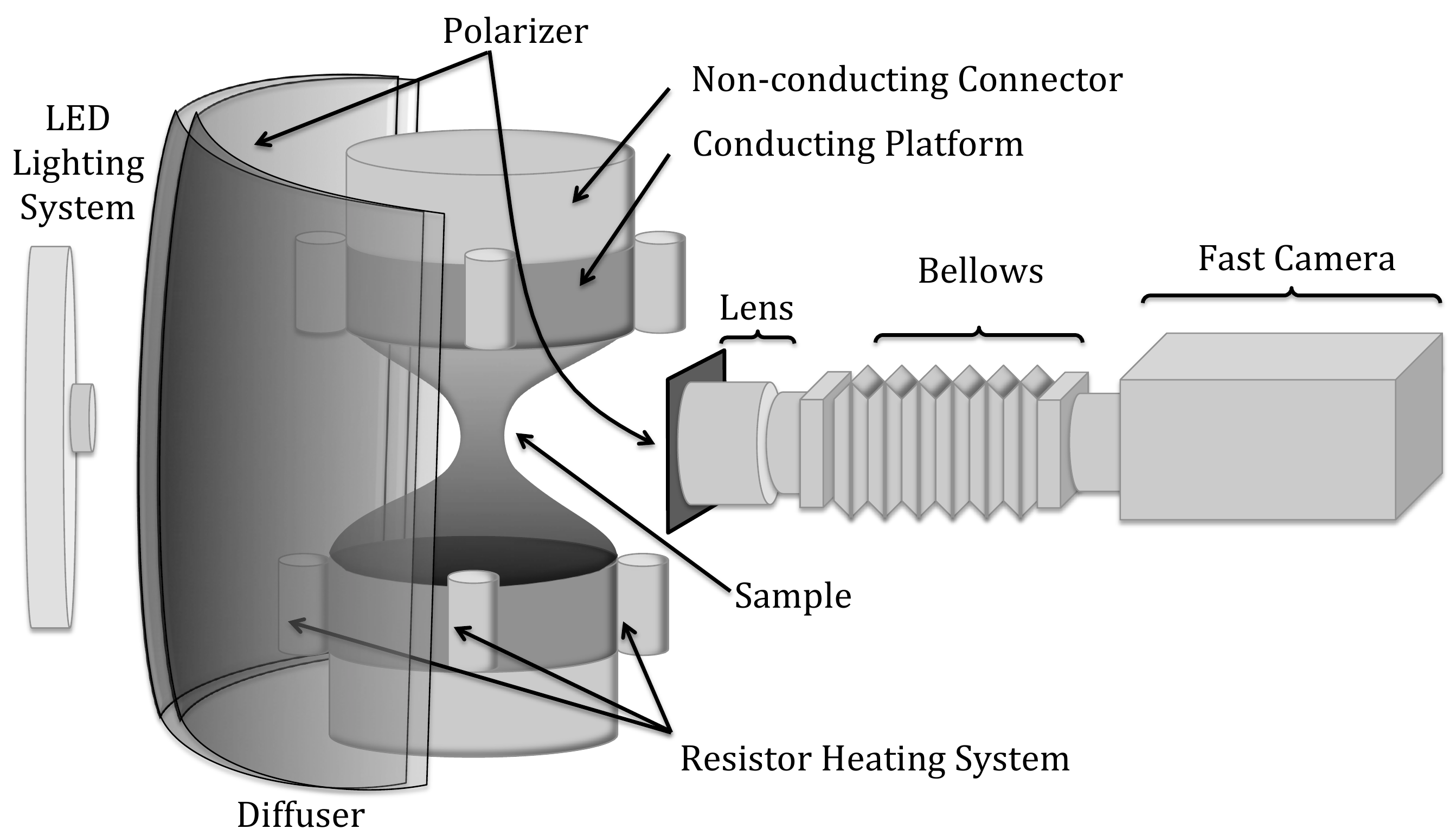}
\end{center}
\caption{Schematic diagram of the apparatus.  The sample fluid is suspended between two heated plates.  The upper plate can be adjusted vertically to induce a narrow fluid bridge suitable for breakup.  A diffuser is wrapped around the bridge in order to increase the area of illumination.  Domain orientation in the sample is probed with polarizers on the camera lens and touching the diffuser.  The bellows facilitate magnification of the point of breakup in the sample.}
\label{fig:setup}
\end{figure}

During the experiments a fluid sample is placed between two conducting metallic plates that can be separated vertically to induce breakup (Figure~\ref{fig:setup}).  Regulated resistors attached to the plates with thermally conductive paste were used to control the temperature. In order to minimize heat loss, Non-conducting plates were used to separate the conducting plates from the rest of the apparatus.  Finally, the resistors selected enabled us to generate temperatures between $\sim 20^\circ $ $\mathrm{C}$ and $60^\circ$ $\mathrm{C}$.  

A fast camera was used to image the breakup at a high frame rate of 22099 fps with a field of view of $128\times 512$ pixels, a medium frame rate of 8213 fps with $512\times 512$ pixels, and a low frame rate of 4796 fps with $800\times 600$ pixels.  A bellows was used to magnify the image at all frame rates so that $1$ pixel was about $3$ $\mathrm{\mu m}$.  The sample was illuminated with a high luminosity LED light and  cross polarizers were used to visualize domain orientation as illustrated in figure \ref{fig:setup}. 

\begin{figure}
\begin{center}
\includegraphics{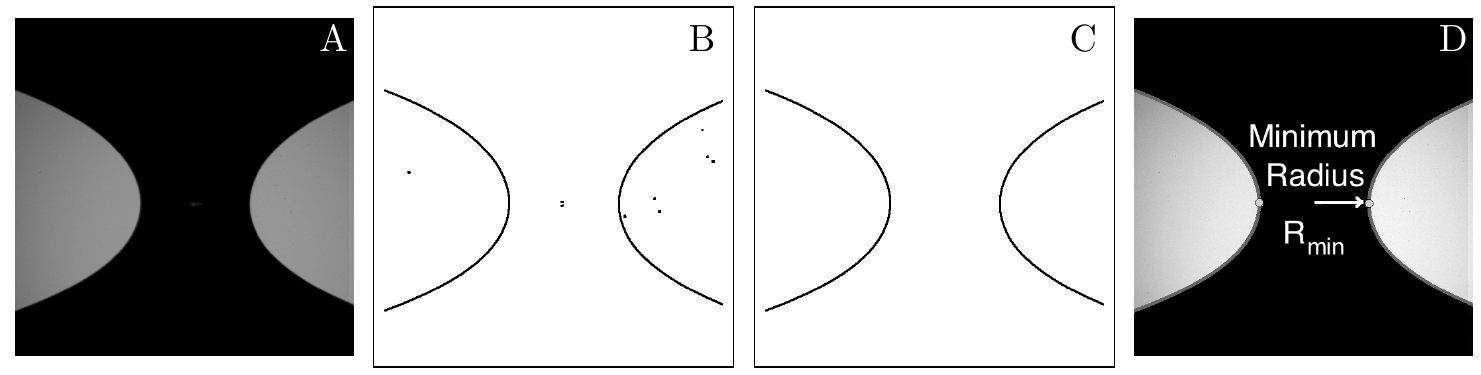}
\end{center}
\caption{The major steps of the analysis program.  Frame A shows the unaltered image captured by the camera.  Frame B illustrates the data points extracted using the Matlab edge identification algorithm.  Using a clustering algorithm we are able to remove the speckle noise resulting from the edge detection program, shown in frame C.  Finally, the profile is fit using a high order polynomial and the minimum radius is measured.  Shown in frame D is the extracted profile overlaid onto the original image along with an illustration of the radius we measure.}
\label{analysis}
\end{figure}

A Matlab code was used to extract the minimum radius as a function of time left to breakup.  The various stages of the analysis are shown in Fig.~\ref{analysis}. To extract the bridge profile from the individual image frames, we used edge detection software (Fig.~\ref{analysis}B) followed by a clustered component analysis that removed noise (Fig.~\ref{analysis}C). The extracted profiles were fit to a high order polynomial from which the minimum radius is extracted at sub-pixel resolution (Fig.~\ref{analysis}D).  To determine $\tau$ we need to identify the time of breakup, which is bounded between two successive frames.  To mediate this uncertainty, we image many breakup events.   By overlaying the points from each movie on top of each other and adjusting the time of the first frame in each movie we have reduced this ambiguity.  

In addition to overlaying points from movies taken at the same frame rate and resolution, we image the breakup using low, medium and high frame rates to probe the small and large $\tau$ behavior.   The data from these regimes are then stitched together using the procedure outlined in the previous paragraph to form a single breakup curve.  To identify scaling behaviors, power law fits to data subsets were made.  In addition, we performed linear fits to the $R_{min}$ versus $\tau$ data in semi-log to identify potential exponential behavior.  We found that the largest source of uncertainty in the fitting arose from the choice of fit range. To estimate this uncertainty we determined how the power law exponents and exponential rates extracted from the fits varied as the range of data was decreased by 30$\%$.  

\section{Results}
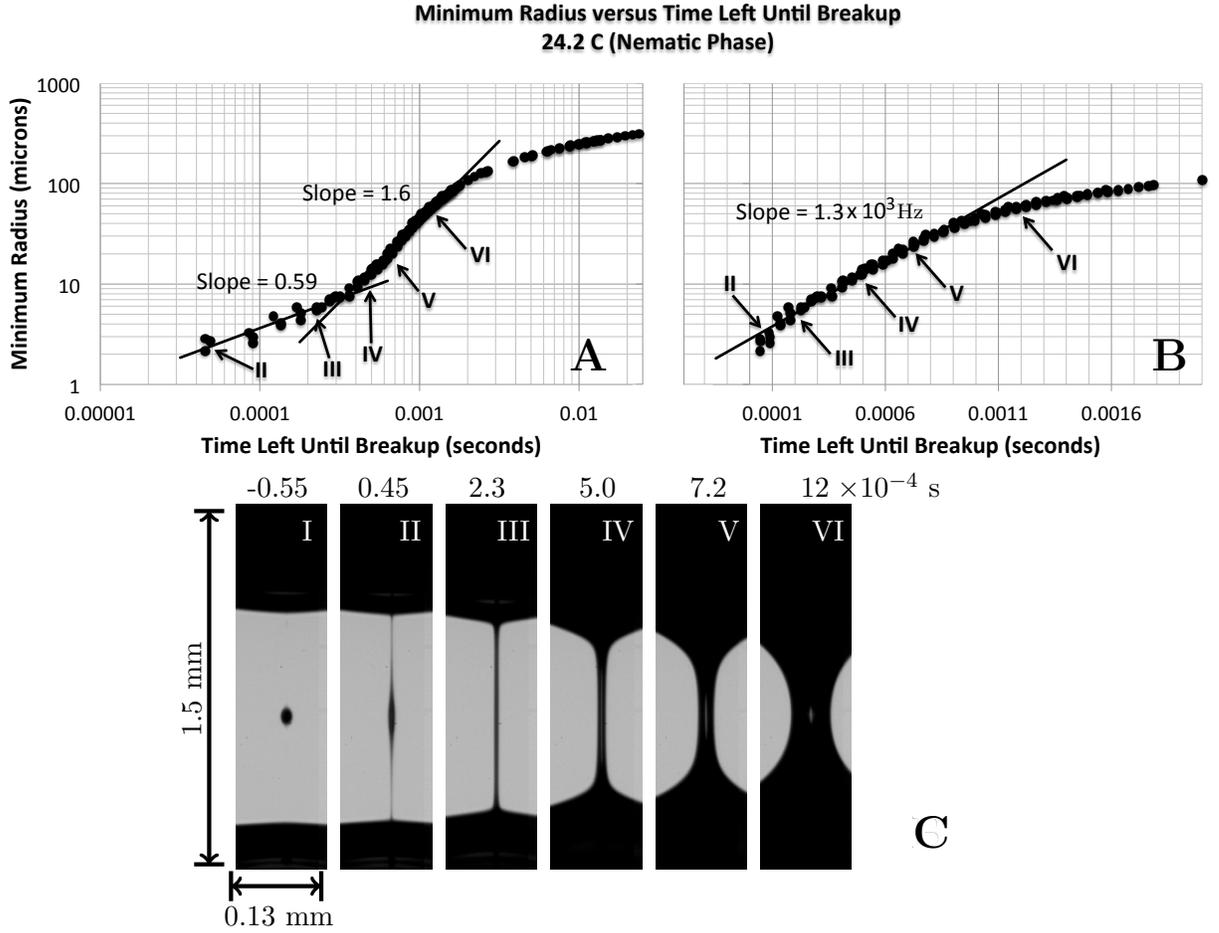
\begin{figure}
\begin{overpic}{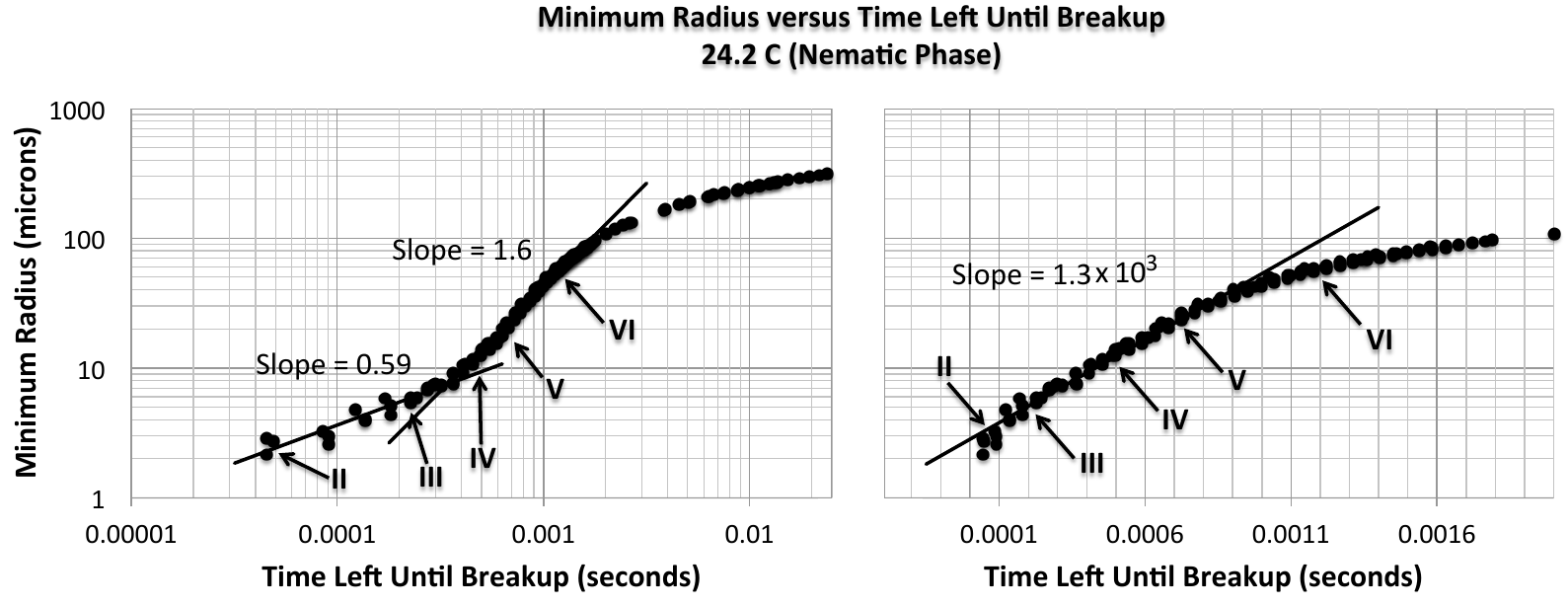}\put(47,8){\bf\Large A}\put(95,8){\bf\Large B}\put(74,20.6){\scriptsize Hz}\end{overpic}\\
\begin{overpic}{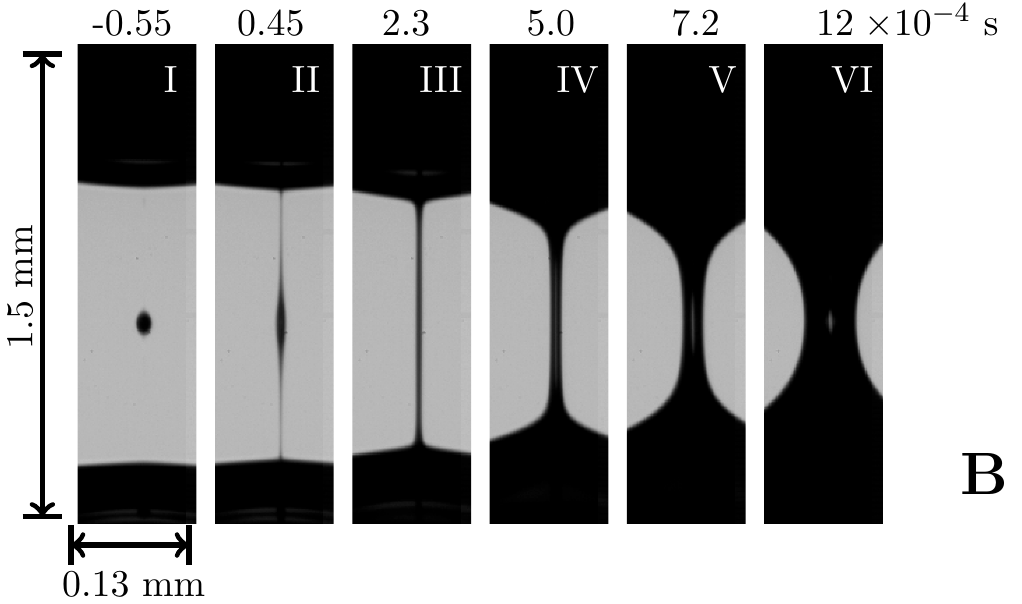}\put(94.1,10.65){\color{white}\bf\Large B}\put(94.1,10.8){\color{white}\bf\Large B}\put(94.3,10.65){\color{white}\bf\Large B}\put(94.3,10.65){\bf\Large C}\end{overpic}
\caption{A\&B. Minimum radius $R_{\mathrm{min}}$ as a function of time left until breakup $\tau$ in the nematic phase in a log-log format (A) and semi-log (B).  Each plot shows data from six movies that have been stitched together.  The ambiguity in the $\tau$ of the first frame was reduced by aligning the six curves overlaid on top of each other.  The behavior for $\tau\gtrsim 0.002$ is a transient regime that depends on the initial geometry more than any final scaling.  In A, we found two power law regimes, one with $R_{min}\sim\tau^{1.6}$ and then $R_{min}\sim\tau^{0.6}$ in the final regime.  In B, we observe a single exponential slope of $1.3\times 10^3$.  CI-CVI. Representative profiles at $T=24.2^\circ$ $\mathrm{C}$.  While we observe a single minimum in profiles CIV-CVI there are two minima in the later profiles CII and CIII.  The breakup terminates in a single satellite droplet.}
\label{nem}
\end{figure}

The dependance of the minimum profile radius for MBBA as a function of time left until breakup was examined at 10 different temperatures that span the isotropic to nematic phase transition. At each temperature we characterized the dependence of the minimum radius on the time left to break up $\tau$ for 6 movies: three at the high frame rate, two at the medium frame rate and one at the low frame rate.  For example, the minimum radius versus $\tau$ for nematic MBBA at $T=24.2^\circ$ $\mathrm{C}$ is shown in Figures~\ref{nem}A\&B.  To test for universal behavior, we plot $R_{min}$ versus $\tau$ on a log-log plot (Figure~\ref{nem}A). We find that the data are consistent with two different power law regimes.  Figures~\ref{nem}C IV,V and VI, show three images from the first regime that is characterized by symmetric profiles with single minima whose radii decreases as $\tau^{1.6}$.  Figures~\ref{nem}C II and III show profiles from the second and final regime that is characterized by two minima whose radii decrease as $\tau^{0.59}$. The fact that the second regime has a lower power law indicates that the dynamics are slowing in the second regime. To test whether such dynamics are exponential in form as is the case in polymeric systems, we plot $R_{min}$ versus $\tau$ in semi-log form (Figure~\ref{nem}B). We find that both the symmetric and asymmetric regimes can be fit with a single exponential form $R_{min}\sim\exp\Big((1.3\times 10^3 \text{ Hz})\tau\Big)$.  Similar behaviors were observed for all 5 bridge profile measurements of MBBA in the nematic phase. 
 
\begin{figure}
\begin{overpic}{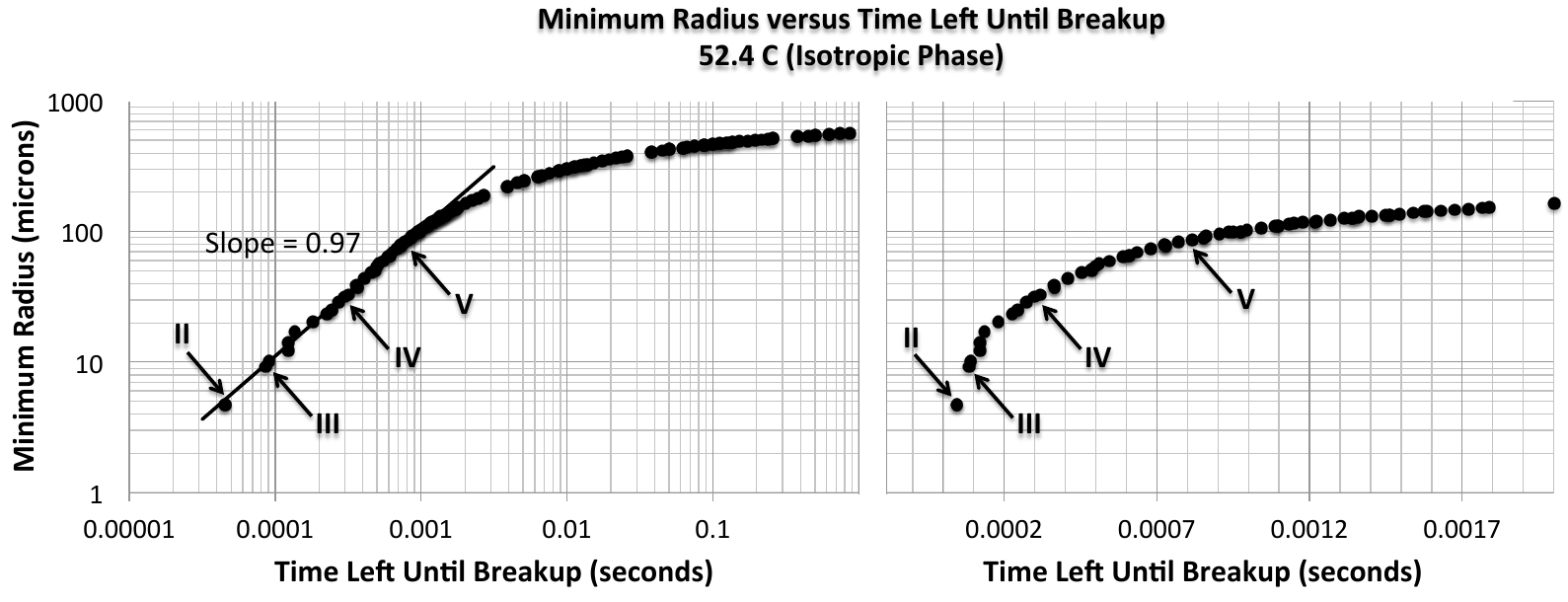}\put(47,8){\bf\Large A}\put(95,8){\bf\Large B}\end{overpic}\\
\begin{overpic}{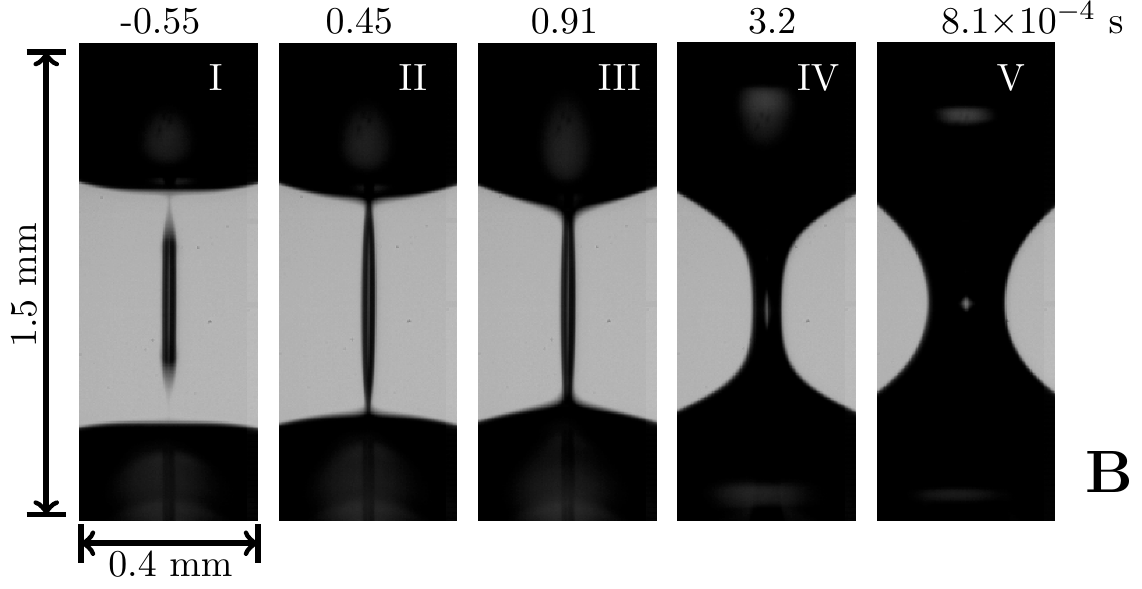}\put(95.1,8.46){\color{white}\bf\Large B}\put(95.1,8.6){\color{white}\bf\Large B}\put(96,8.46){\color{white}\bf\Large B}\put(95.1,8.46){\bf\Large C}\end{overpic}
\caption{A\&B. Minimum radius $R_{\mathrm{min}}$ as a function of time left until breakup $\tau$ in the isotropic phase in a log-log format (A) and semi-log (B).  Each plot shows data from six movies that have been stitched together.  The ambiguity in the $\tau$ of the first frame was reduced in aligning the six curves overlaid on top of each other.  The behavior for $\tau\gtrsim 0.002$ is a transient regime that depends on the initial geometry more than any final scaling.  In A, we found a single power law regime with $R_{min}\sim\tau$.  In B, we observe that an exponential model does not well describe the data.  CI-CVI.  Representative profiles at $T=24.2^\circ$ $\mathrm{C}$.  CII and CIII show the initial symmetric behavior while CIV and CIV show the final asymmetric breakup.}
\label{iso}
\end{figure}

At higher temperatures where MBBA was determined to be isotropic the breakup behavior was markedly different than breakup in the nematic regime. For example, the minimum radius versus $\tau$ for isotropic MBBA at $T=52.4^\circ$ $\mathrm{C}$ is shown in Figures~\ref{iso}A\&B.   From Figure~\ref{iso}A, we conclude that in contrast to the nematic breakup, the power law behavior consists of a single asymptotic exponent.   As seen in Figures~\ref{iso}C II-V for both the initial symmetric (\ref{iso}CII, III) and final asymmetric (\ref{iso}CIV, V) regimes the minimum radius decreases as $\tau^{0.97}$. This exponent and the final breakup profiles are consistent with the behavior observed in drop breakup in the viscous-inertial asymptotic regime of Newtonian fluids \cite{Shi1994}.  We observe that unlike for the nematic regime shown in Figure~\ref{nem}B, the data are inconsistent with an exponential dependance of $R_{min}$ on $\tau$ in the isotropic regime (Figure~\ref{iso}B).

\begin{figure}
\begin{center}
\includegraphics[scale=0.5]{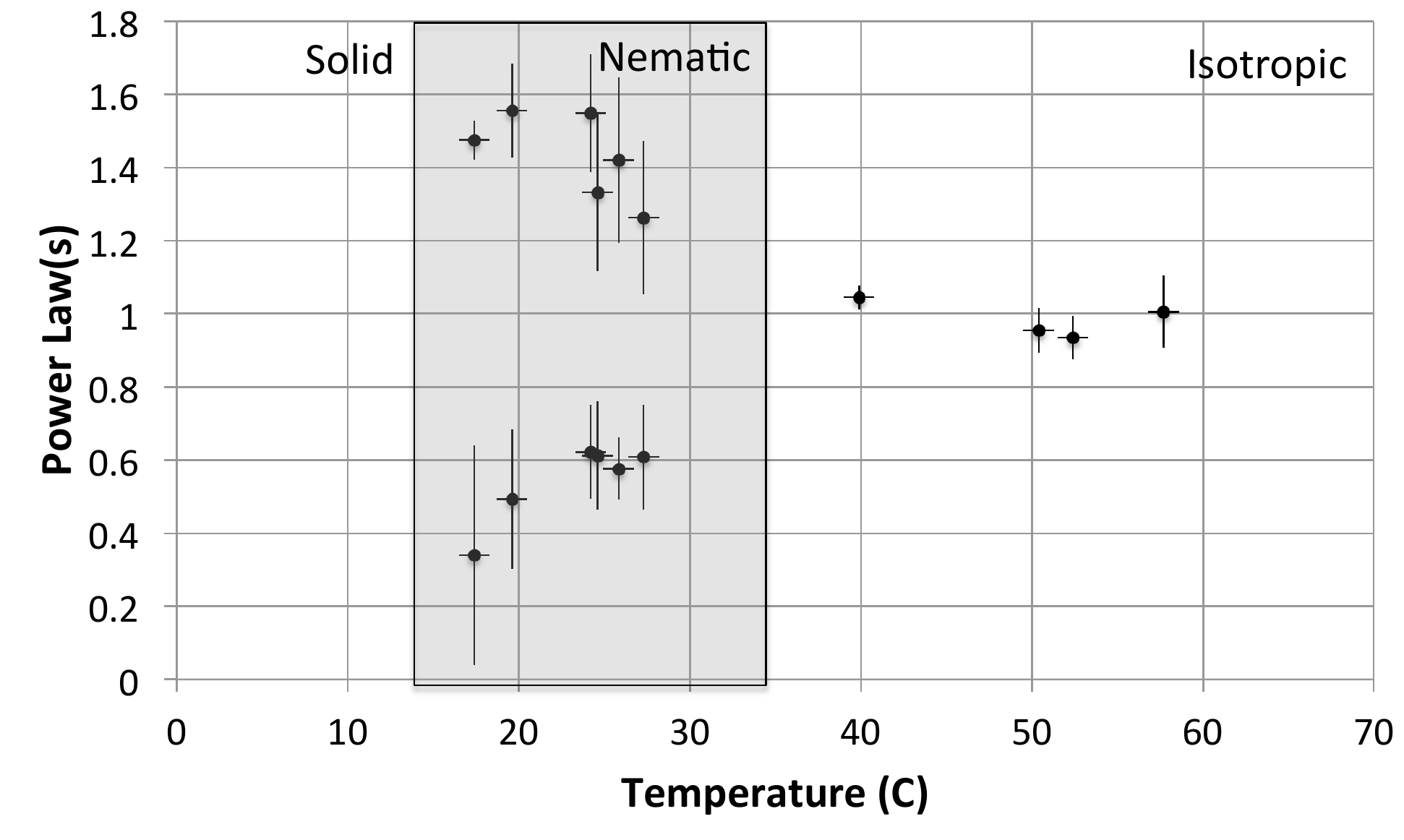}
\end{center}
\vspace{-7mm}
\caption{Shown here are the power laws extracted from the breakup profiles of six movies for each temperature.  In the nematic regime (gray shaded region), there are two different power laws.  The larger power laws correspond to the initial symmetric behavior where the bridge has a single minimum (e.g. Fig.~\ref{nem}C IV,V and VI) while the lower branch documents the behavior in the final asymmetric regime where the bridge has two minima (e.g. Fig.~\ref{nem}C II and III).  Finally, at temperatures above $35^\circ$ C where birefringence data shows MBBA is isotropic, we find a single power law exponent describes both the symmetric and asymmetric regimes.}
\label{res}
\end{figure}

\begin{figure}
\begin{center}
\includegraphics[scale=0.68]{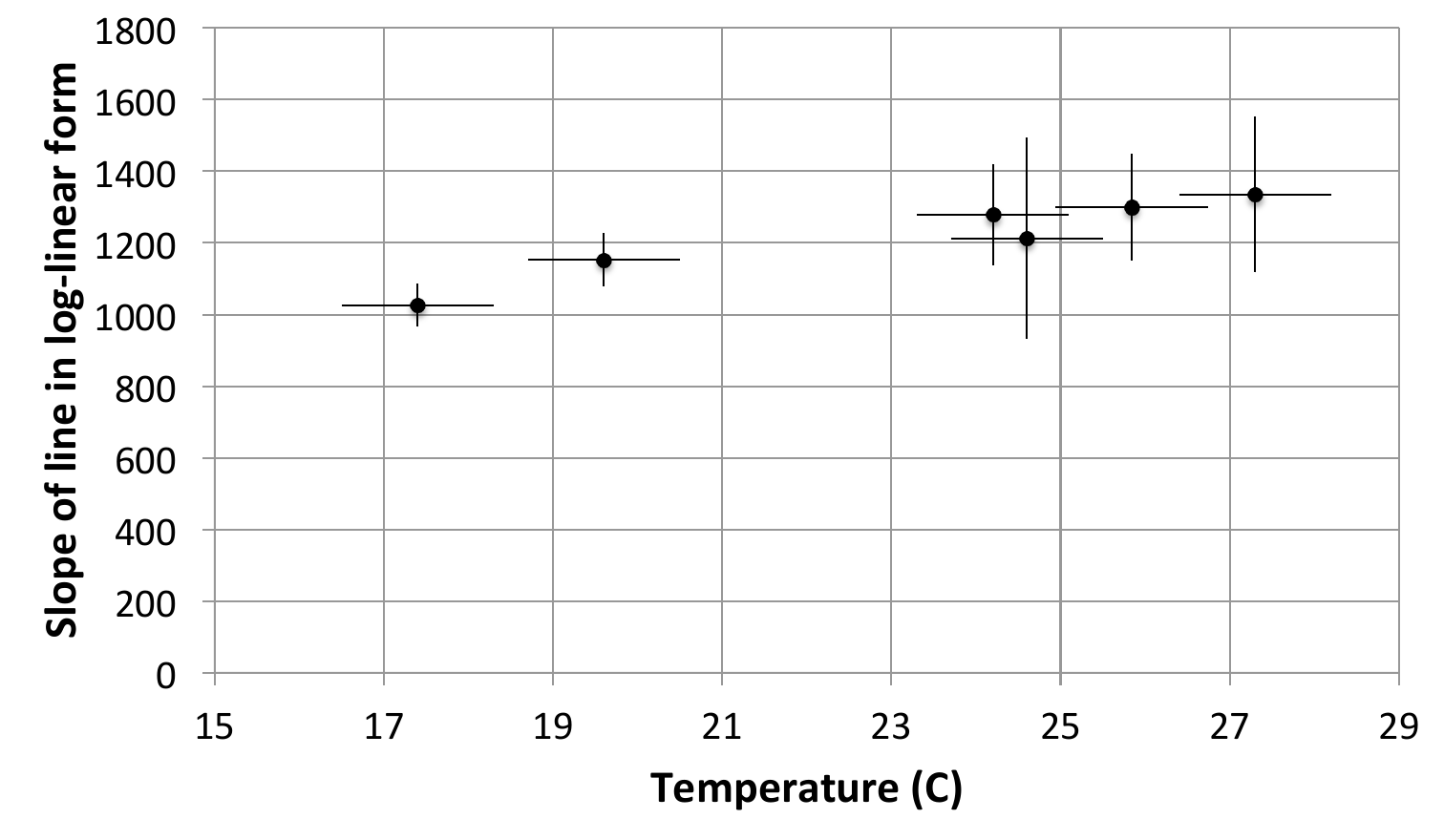}
\end{center}
\vspace{-7mm}
\caption{Shown here are the slopes (in Hz) of the exponential models fit to $R_{min}$ versus $\tau$ in the semi-log representation, i.e. the coefficient of the linear term in $\log(R_{min})=c_1\log\tau+c_0$.  We find that this fit well-describes the data in the nematic regime (c.f. Fig.~\ref{nem}B) but does not model the data well in the isotropic regime (c.f. Fig.~\ref{iso}B) and so the plot is redacted to show only the nematic behavior.  For each temperature, the slope was extracted from the breakup of six movies.}
\label{res2}
\end{figure}

We summarize the temperature dependence of the power-law exponents and exponential rates in Figures~\ref{res} and~\ref{res2}, respectively.  We find that the power law behavior for all ten temperatures in both the nematic and isotropic regimes shows no significant temperature dependence (Fig.~\ref{res}).  Averaging over the different measurements we find that in the nematic phase, the first breakup regime is characterized by a symmetric profile with a single minimum whose radius decreases as $\tau^{1.51 \pm 0.06}$. The second and final regime is characterized by two minima whose radii decrease as $\tau^{0.52 \pm 0.11}$.  For the isotropic phase in both the initial symmetric and final asymmetric regimes the minimum radius decreases as $\tau^{1.03 \pm 0.04}$. Finally, we note that in both the isotropic and nematic regimes, the breakup is followed by the formation of a large satellite droplet, seen in Figures~\ref{nem}CI and Figure~\ref{iso}CI.  Similarly, the fitted exponential rates show no significant temperature dependence and are characterized by $R_{min}\sim\exp\Big((1.2\pm 0.1\times 10^3\text{ Hz})\tau\Big)$ (Fig.~\ref{res2}). 

\begin{figure}
\begin{center}
\begin{overpic}[scale=0.2]{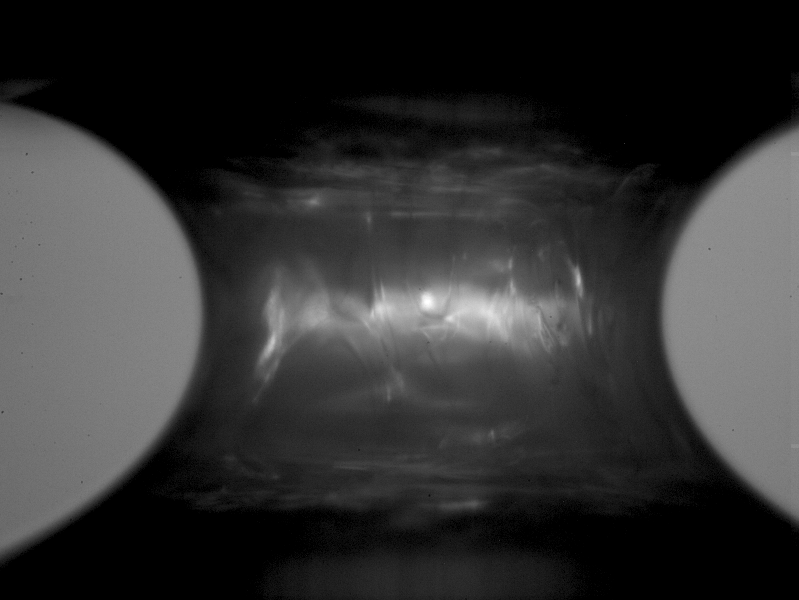}\put(85,65){\color{white}A}\end{overpic}\hspace{5mm}\begin{overpic}[scale=0.2]{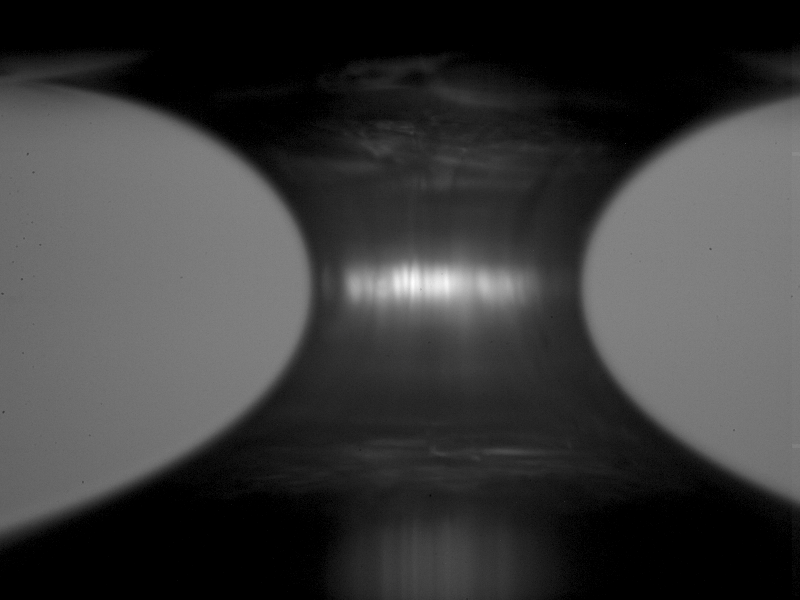}\put(85,65){\color{white}B}\end{overpic}
\end{center}
\caption{Breakup under cross polarizers.  To improve the image, we artificially lightened the background (uniform light gray).  We note however that the features within the drops were not altered. Profile A was imaged at $5.9$ seconds prior to breakup.  At $0.2$ seconds before breakup (profile B), one can find that the domains start aligning well before the final regime is reached.}
\label{xp}
\end{figure}

Since MBBA is translucent, we are able to probe the breakup behavior under cross polarizers.  Figure~\ref{xp} shows sample profiles in the nematic regime (room temperature) at different times prior to breakup.  In Figure~\ref{xp}A, we find that the structural domains in the bridge are randomly oriented.  In contrast, in Figure~\ref{xp}B, we observe large scale alignment of the domains to a direction that is parallel with the surface of the fluid. This alignment initiates early in the breakup process. Therefore, it will be important to take into account the rheology arising from domain realignment in order to determine the forcing terms that govern the breakup dynamics.  

Unfortunately, we could not examine the profiles under cross polarizers for small $\tau$.  Given the current camera sensitivity, the main limitation was in the intensity of light that illuminated the sample.  Since MBBA is translucent and not transparent, a large backlight intensity is required, so to produce Figure~\ref{xp} we had to compensate by increasing the exposure time at the cost of a lower frame rate.  One could try to add more LEDs at a higher intensity, but this is nontrivial because order of magnitude estimates show that we need to increase the intensity by about a factor of 100 in order to raise the frame rate enough to see the asymptotic breakup regimes.  A possible method for achieving this intensity is with a laser.  For example, with phase-modulated flow birefringence it should be possible to determine the orientation angle at a point~\cite{Frattini1984,Rothstein2003,Rothstein2002}.  However, such a method will only report on the local orientation. Therefore, obtaining more comprehensive information about domain behaviors and alignments would require scanning at multiple positions in the bridge and accounting for the deflections of the incident beams as the bridge breaks.


\section{Conclusions}

The tabulated results in Figures~\ref{res} and~\ref{res2} confirm that the asymptotic breakup behavior of MBBA depends on the liquid crystal phase.  In the universal interpretation, the power law changes as the fluid transitions between the isotropic and nematic phases.  As was noted in~\cite{Porter2012}, the final power law regime in the nematic phase is consistent with the 2/3 power law expected in the inertial Newtonian regime.  However, qualitative differences in the profile suggest very different radial and axial scalings~\cite{Day1998} and therefore rule out this scaling relation.  In the nematic phase, the asymptotic behavior is described by two power laws which correspond to qualitative changes observed in the droplet profile.  In contrast, in the isotropic regime, the data is well modeled by a single exponent for both the symmetric and asymmetric final breakup profiles.  Furthermore, this behavior is consistent with observed drop breakup in the viscous-inertial asymptotic regime of Newtonian fluids. Bulk rheology measurements indicate Newtonian behavior with a constant viscosity in both the isotropic and anisotropic regimes for shear rates of up to 2 kHz~\cite{Negita1995}. Nevertheless, strain thinning behavior has been shown to be important for determining breakup similarity solutions observed in smectic liquid crystals~\cite{Savage2010}. One possible resolution to these contrasting observations is that the breakup flows may be exploring substantially higher shear rates where non-Newtonian rheologies might become important.   

In the exponential interpretation, the temperature dependance is clear from the fact that the asymptotic behavior has a different relationship between $R_{min}$ and $\tau$ across the phase transition.  The universal scaling in the isotropic regime is consistent with expectations.   Lacking the additional structure from long range molecular orientations, MBBA in the isotropic regime resembles a Newtonian fluid in its asymptotic behavior.   In contrast,  the asymptotic behavior of MBBA in the nematic phase is well described by a single exponential curve.  Furthermore, this relationship seems to hold across various temperatures, as summarized in Fig.~\ref{res2}.  The exponential model has a clear motivation that can be understood in an analogy to polymeric systems.  Studies have shown that the power law dependance of $R_{min}$ on $\tau$ in a Newtonian fluid can change abruptly to an exponential relationship when polymers are added~\cite{Amarouchene2001,Roche2011,Anna2001}.  This transition is due to the elasticity of the polymer molecules.  The increasing energy required to stretch the polymers is realized as an increase in the effective extensional viscosity~\cite{Eggers2008}.  In a similar way, there is an increasing energy cost in liquid crystal systems.  With nematic ordering, there is a preferred long range order for the molecules within a liquid crystal domain.  Near breakup, topological constraints may require the molecules to rearrange. This rearrangement can lead to elastic stresses and slower thinning.  

While the current dataset cannot differentiate between the power law scaling and the exponential model, the later has a clear motivation.  In the universal interpretation, the measurements for the the nematic and isotropic regimes are consistent with the corresponding scalings found in the same phases of 5CB and 8CB~\cite{Porter2012,Savage2010}, indicating that similar dynamics govern all three systems.  Therefore, if further studies reveal that exponential slowing is a more appropriate model for the asymptotic behavior of these systems, we expect this conclusion to hold more generally.  Higher precision measurements of the asymptotic breakup behavior coupled with further analysis of domain orientation will elucidate the underlying power law or  exponential dependance.  In addition, such studies will be able to pinpoint the mechanism that is driving the observed flows.

In addition to bulk rheology measurements of MBBA, studies have been conducted to investigate the average orientation of the nematic directors with respect to the flow direction.   We have already noted that for nematic fluids with domain sizes on the order of the droplet dimensions this orientation of nematic directors influences the breakup dynamics in addition to the thermodynamic phase~\cite{Zhou2010,Cheong2004,Goyal2008,Lekkerkerker2012}.   In our analysis of breakup in MBBA, where the domain size is much smaller than the initial droplet size, we have probed the dependance of scaling only on the phase.  There are several methods for extending the study to additionally probe the dependance on director orientation.  For example, there is a known relationship between temperature and preferred director angle with respect to the flow direction~\cite{Gahwiller1972,Meiboom1973,Kobinata1981,Kneppe1982}.  If the measurement uncertainties can be reduced  it may be possible to elucidate how director alignment alters asymptotic breakup.    Additionally, external electric fields have been related to director orientation~\cite{Negita1995}, and could therefore also be used as a control knob for tuning alignment and possibly breakup behavior. Having established a clear link between the nematic liquid crystal phase and the measured behavior for the evolution of the bridge profile, it will be important to identify and develop handles on the specific effects of orientation to understand and ultimately control this type of breakup behavior.

One handle that has been utilized for the first time in this experiment is the in situ imaging of a nematic liquid crystal breakup under cross polarizers.  The analogous measurements for a smectic liquid crystal~\cite{Savage2010} provided insight into  the liquid crystal orientation and our measurement has initiated this study for the nematic liquid crystal domain orientation in MBBA.   Future work in this direction will need to find ways for increasing the intensity for illuminating the sample, but ideas already exist to begin this research.   With the added complexity of nematic liquid crystals, traditional techniques must be extended in order to fully understand the breakup dynamics.   However, with tools like in situ imaging under cross polarizers, we can begin to probe the deeper relationship between small scale nematic liquid crystal structure and bulk flows during droplet breakup.

\section{Acknowledgements}

We thank Jim Sethna for discussions that brought the exponential interpretation to our attention.  In addition, this work was partially supported by Proctor and Gamble and we thank our collaborators there Patrick Spicer and Marco Caggioni.   This work would not have been possible without the extensive support of the entire Cohen Lab at Cornell.  In particular, we would like to thank John Mergo for helping construct the apparatus, Jeanette Nguyen for her help with procuring the lighting system, Brian Leahy for help with the minimum radius code and Tsevi Beatus for many useful discussions and overall support.   

\clearpage
\newpage

\bibliographystyle{plainnat}
\bibliography{myrefs.bib}{}

\end{document}